\RequirePackage[l2tabu, orthodox]{nag}
\documentclass[twocolumn,amsmath,amssymb,aps,showkeys,nofootinbib]{revtex4-2}

\usepackage{graphicx}
\graphicspath{{figures/}}
\usepackage{dcolumn}
\usepackage{bm}
\usepackage[hidelinks]{hyperref}

\usepackage{titlesec}
\titleformat{\section}[hang]{\normalfont\bfseries}{\thesection.}{0.5em}{}[]
\titlespacing{\section}{0pt}{*2}{*1}

\begin{document}

\title{A generating function perspective on the transmission forest}
\author{Niket Thakkar}
\email{niket.thakkar@gatesfoundation.org}
\homepage[\\]{https://github.com/NThakkar-IDM/covid_and_stat_mech}
\author{Mike Famulare}
\affiliation{%
The Institute for Disease Modeling\\
Global Health | Bill $\&$ Melinda Gates Foundation\\
Seattle, Washington 98109
}%
\date{\today}

\begin{abstract}
In a previous paper, we showed that a compartmental stochastic process model of SARS-CoV-2 transmission could be fit to time series data and then reinterpreted as a collection of interacting branching processes drawn from a dynamic degree distribution. We called this reinterpretation a transmission forest. This paper builds on that idea. Specifically, leveraging generating function methods from analytic combinatorics, we develop a theory describing the transmission forest's properties, allowing us to show for example that transmission tree interactions fade with increasing disease prevalence. We then validate the theory by computing forest statistics, like the tree survival function, which we compare to estimates based on the sampling method developed previously. The accuracy and flexibility of the analytic approach is clear, and it allows us to comment on multi-scale features of more general transmission processes. 
\end{abstract}
\maketitle

\section{Epidemiology and graph theory}
Transmission trees are the basic graphical unit of an epidemiology. Or, said differently, if we knew the full transmission tree, it would shed light on a range of questions -- characteristics assigned to nodes would inform our understanding of risk, features of edges would teach us about transmission mechanisms, and geometric changes would help us estimate the effects of interventions. 

But of course the basic epidemiological problem is that we can't measure transmission trees directly. Even in perfectly observed, closed populations, assigning edges between nodes is only possible inferentially when transmission events are sufficiently staggered. In that situation, the gold standard is outbreak investigations conducted by specialists, where interviews and follow ups are used to construct a plausible subtree of the full transmission tree \cite{arunkumar2019outbreak}. As a result, even in the best case, this approach cannot scale to a full epidemiological system. 

Another, modern, alternative path forward comes from work on viral phylogenetics. In that case, genetic sequences of sampled viruses can be arranged into a phylogenetic tree, and features of that tree can be used to infer features of the associated transmission trees \cite{grenfell2004unifying}. This is easier in theory than in practice. The phylogentic tree is a product of epidemiological and evolutionary processes, the latter of which can be very complex and system specific, and separating signals in general is a challenge. 

Within this context, in a paper last year \cite{thakkar2022covid}, we explored a third perspective. We demonstrated that volatility in Washington's COVID-19 epidemiological curves contains information on the underlying transmission trees' degree distribution. Once specified, that time-varying distribution could be used to grow a set of interacting trees, which we called a transmission forest, and while those trees lack the individual-level resolution of conventionally estimated transmission trees, they were shown to be predictive of observations from outbreak investigations and phylogenetics. 

This step towards harmony between often discordant views of a cryptic epidemic \cite{bedford2020cryptic}, that is the time series, phylogeny, and outbreak investigations, speaks to the transmission tree as a fundamental structure for organizing epidemiological information. Transmission tree inference is an ambitious goal, but we stand to learn a lot working towards it.

Along those lines, this paper contributes to the broader project of statistically characterizing transmission trees using more readily available epidemiological data. Our concrete goal is to more completely and rigorously define the transmission forest, and in some sense this paper is a mathematical supplement to the original \cite{thakkar2022covid}. That said, more than clarifying the work from last year, the formalism we develop here offers a tractable connection between individual-level behavior, the resulting tree structures, and emergent forests, leading to broadly applicable multi-scale insights into transmission processes. 

\section{The transmission forest model}

\begin{figure*}
\centering\includegraphics[width=\linewidth]{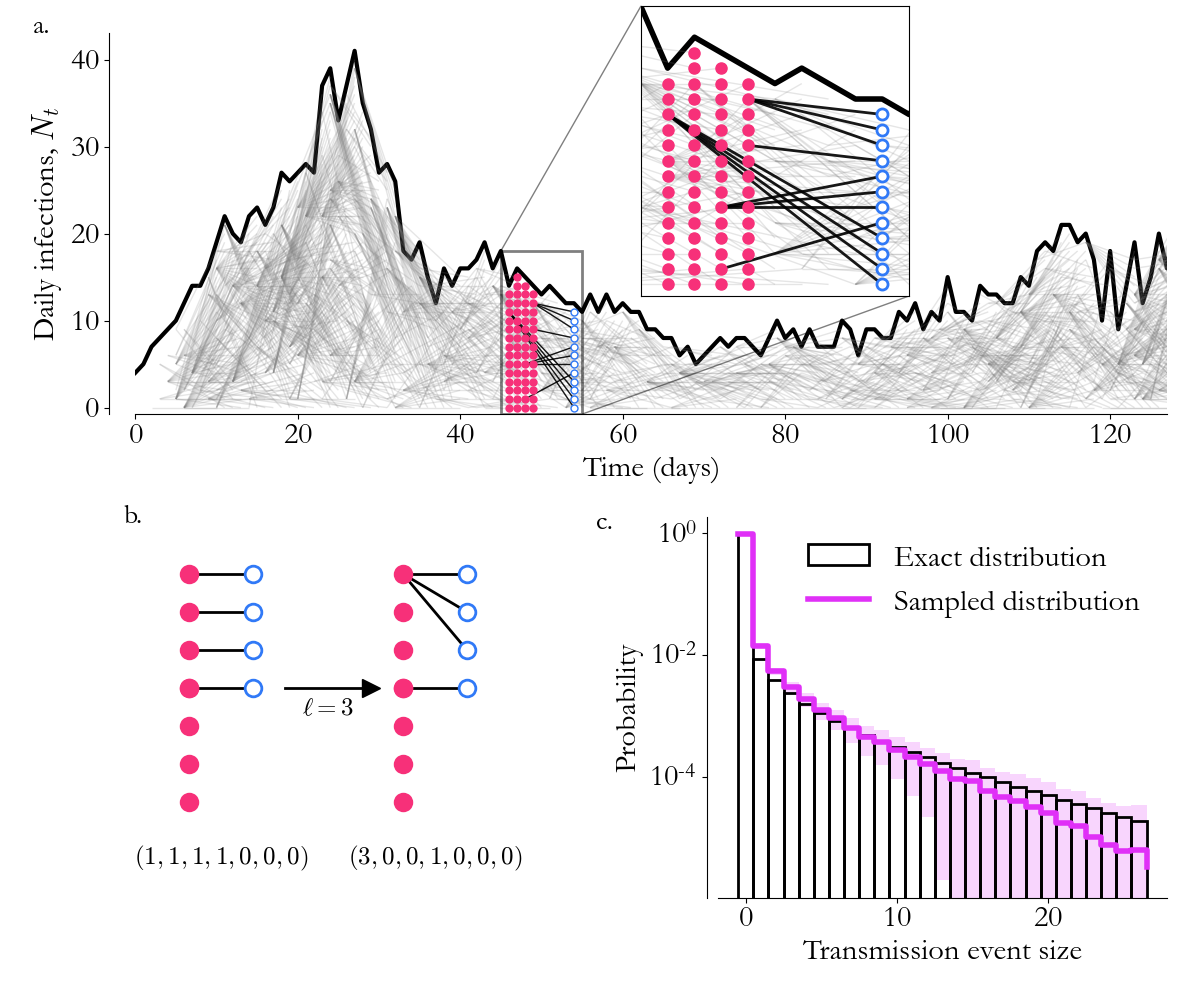}
\caption{Sampling the forest. (a) For a given $N_t$ trajectory (black curve) a transmission forest is a random graph (edges in grey) constrained to sum to $N_t$. A given day's transmission (inset) between infectious parents (red dots) and their newly infected children (blue circles) demonstrates how trees compete for nodes. (b) We can sample under the sum constraint by creating transmission constellations from a fixed number of edges. (c) The sampling approach (10k samples, mean in purple, 95$\%$ CI shaded) is able to efficiently explore tail-events in the skewed degree distributions (black bars) underlying transmission.}
\label{fig:sampling}
\end{figure*}

We start by defining the transmission forest mathematically and in that way encapsulating many of the previous paper's main ideas. Consider a discrete time, stochastic disease transmission process between susceptible individuals, $S_t$, and infectious individuals, $I_t$. In a closed population, we have
\begin{align}
    S_t &= S_{t-1} -  N_{t-1} \label{eq:SIR} \\
    E_t &= \left(1 - \frac{1}{d_E}\right)E_{t-1} + N_{t-1}\\
    I_t &= \left(1 - \frac{1}{d_I}\right)I_{t-1} + \frac{1}{d_E}E_{t-1},
\end{align}
where newly infected individuals, $N_t$, become exposed but not infectious, $E_t$, for $d_E$ days on average before becoming infectious for $d_I$ days on average. On the one hand, following classical models \cite{anderson1992infectious}, we can write $N_t$ as
\begin{equation}
    N_t = (\beta_t\varepsilon_t)S_tI_t, \label{eq:sir_view}
\end{equation}
with transmission rate $\beta_t$ and log-normally distributed volatility $\varepsilon_t$, capturing the idea that new infections come from a random fraction of interacting pairs. Equivalently, inspired by branching processes \cite{fewster2014stochastic}, we also have
\begin{equation}
    N_t = \sum_{i=1}^{I_t} T_{it}, \label{eq:bp_view}
\end{equation}
capturing the idea that new infections are the sum total of realized daily transmission events, $T_{it}$, across infectious individuals indexed by $i$. We model transmission events as independent and identically distributed with 
\begin{equation}
    T_t \sim \text{NegBin}\left(\mu_t,k_t\right), \label{eq:NB}
\end{equation}
which is approximately entropy maximizing \cite{jaynes1957information} when the mean, $\mu_t$, and over-dispersion, $k_t$, are set to maintain Eq. \ref{eq:bp_view} up to second order moments \cite{thakkar2022covid}.

The transmission forest is a random graph drawn from this model. We visualize a sample in Fig. \ref{fig:sampling}a. For a given $N_t$ trajectory (black curve), nodes are placed in daily columns and directed edges are drawn between nodes on day $t$ (blue circles) and their infectious parents between days $t-d_E$ and $t-d_E-d_I$ (red dots) according to Eq. \ref{eq:NB}. The process is repeated for all $t$ to fill the graph (grey lines, where we've suppressed the nodes for visual clarity).\footnote{This definition makes it clear that we've taken $d_E$ and $d_I$ to be deterministic across nodes. Strictly speaking, that approximation is optional. In both the sampling approach and the theory, latent and infectious durations can be arbitrarily distributed, but this simplifying assumption is in keeping with the previous paper, and it helps us focus the work.} In the early part of the trajectory, where $t < d_E + d_I$, nodes have no parents, making the graph a necessarily disjoint collection of transmission trees, inspiring the name.

Model fitting to observed time series is discussed in detail in our previous paper -- here we assume the parameters $\beta_t$ and $\varepsilon_t$ as well as the process's initial conditions are known. Throughout this paper, we use the same model fit to COVID-19 data from Washington from January 2020 to March 2021 to illustrate the theory's application. 

\section{Sampling a forest}

To generate sample forests as in Fig. \ref{fig:sampling}a, we draw daily transmission graphs, shown in the inset, and stitch them together over time. The sampling approach gives scaffolding to the theory below, and we use it to validate results throughout, so it's worth a brief aside.

$I_t$ and $N_t$ trajectories can be drawn from Eqs. \ref{eq:SIR}--\ref{eq:sir_view} using standard normal samples. Then, for a given trajectory (rounded to the nearest integers), the number of nodes in the graph is fixed, and as a result edges drawn from Eq. \ref{eq:NB} are constrained to satisfy Eq. \ref{eq:bp_view} and are no longer independent. In other words, that trees compete for nodes gives rise to their interactions in the forest.

We might consider a rejection sampler in this case, essentially drawing $I_t$ negative binomial samples repeatedly until they sum to $N_t$ for every $t$, but that approach is unusably inefficient. With a fixed trajectory, the right-hand-side of Eq. \ref{eq:bp_view} is asymptotically Gaussian by the central limit theorem, implying that on average $\mathcal{O}(I_t\sqrt{I_t\text{V}[T_t]})$ samples have to be drawn per time-step to create a single forest sample. For the Washington COVID-19 model we're working with, with a roughly 400 day time series, that translates to over 1 billion negative binomial samples to draw one forest.

Fig. \ref{fig:sampling}b illustrates a way forward. The daily transmission graph is a collection of star graphs, which we call a constellation, and can be represented as a length $I_t$ tuple of infectious node degrees constrained to sum to $N_t$. Initializing a constellation as 
\begin{align*}
    c = (1, 1, 1, ..., 1, 0, 0, 0, ..., 0),
\end{align*}
a tuple of $N_t$ ones and $I_t-N_t$ zeros, we start at $c$'s first entry ($i=0$, representing edges assigned to the first red dot) by defining the rolling sum $C = \sum_{j=0}^{i-1} c_j$ and drawing a sample, $\ell$, from
\begin{align*}
    p(T_t = \ell|N_t,C) = \frac{p(T_t=\ell)}{1 - \sum_{m > N_t-C} p(T_t=m)},
\end{align*}
the conditionally renormalized degree distribution. We then set $c_i = \ell$, set the next $\ell-1$ entries to 0 (taking edges from next red dots), and then repeat the process with $i\rightarrow i+\ell$ until $C = N_t$. In Fig. \ref{fig:sampling}b, this is illustrated for $\ell = 3$, and in Fig. \ref{fig:sampling}c, we verify that the approach can maintain the skewed target distributions we have in mind. Finally, completing the forest requires us to link stars over time. We do this in a simple way, shuffling $c$ and assigning the resulting events to infectious nodes in order, without regard to past assignments. 

This edge rewiring approach requires $\mathcal{O}(N_t)$ negative binomial samples and is, as a result, efficient enough to be performed every day for a given model trajectory. In the Washington example, it leads to a 3 orders of magnitude speed up over the rejection method, allowing us to sample large sets of forests and compute statistics empirically. 

\section{Tree interactions are weak}

Setting aside the empirical approach for now, our goal is to calculate probabilities, like the distribution of tree sizes or their chance of extinction over time. The next sections are the core of this paper, developing a generating function \cite{flajolet2009analytic} approach for modeling forest growth. 

The sampling algorithm above motivates an overarching strategy, first characterizing the possible daily transmission constellations (Fig. \ref{fig:sampling}a inset) and then considering the process of assigning stars to trees. With that in mind, the negative binomial distribution in Eq. \ref{eq:NB} can be represented as a probability generating function \cite{miller2018primer},
\begin{align*}
    f_{T_t}(z) &= p(T_t=0) + p(T_t=1)z + p(T_t=2)z^2 + ...\\
    &= \sum_{\ell \geq 0} \binom{k_t + \ell - 1}{\ell} (p_t z)^{\ell}(1-p_t)^{k_t}\\
    &= \left(\frac{1-p_t}{1-p_tz}\right)^{k_t}
\end{align*}
where $p_t \equiv \mu_t/(\mu_t+k_t)$ and $z$ is an arbitrary complex number. Then, for a given day's $I_t$, constellations drawn from independent samples can be organized by the total number of edges in a generating function product,
\begin{align*}
    C_t(z) &= f_{T_t}(z)^{I_t} = \left(\frac{1-p_t}{1-p_tz}\right)^{I_t k_t} \\
    &= \sum_{n \geq 0} p(n \text{ edges})z^n.
\end{align*}
As a first illustration of a probability calculation, using $[z^n]$ to signify extracting the $z^n$'s coefficient from a formal power series, Newton's binomial theorem implies
\begin{equation}
    [z^{N_t}]C_t(z) = \binom{I_tk_t + N_t - 1}{N_t}p_t^{N_t}(1-p_t)^{I_tk_t}\label{eq:denom}
\end{equation}
is the probability of satisfying Eq. \ref{eq:bp_view} for a fixed $N_t$ over all possible constellations with $I_t$ infectious nodes.  

Eq. \ref{eq:denom} represents the subset of graphs consistent with the trajectory constraints, that is the $I_t$ and $N_t$ node populations. To get a more detailed view of constellation structure, we can calculate the probability of an infectious individual infecting $m$ new people across constrained graphs. Introducing arbitrary complex number $u$ to highlight \cite{flajolet2009analytic} an $m$-event,
\begin{align*}
    &C_{t,m}(z,u) = \left(p(T_t = m)uz^m + \sum_{\ell \neq m} p(T_t=\ell)z^{\ell} \right)^{I_t}\\
    &= (1-p_t)^{k_tI_t}\\
    &\quad\times\left[\binom{k_t+m-1}{m}(u-1)(p_tz)^m+ (1-p_tz)^{-k_t}\right]^{I_t},
\end{align*}
gives a path towards calculating the analog of Eq. \ref{eq:NB} with tree interactions. Note that $C_{t,m}(z,1)=C_t(z)$ for any $m$, as required. Then, the expected number of $m$-events in day $t$'s constellation, $n_m$, is
\begin{equation*}
    \text{E}[n_m | N_t, I_t] = \frac{[z^{N_t}]\frac{\partial}{\partial u}C_{t,m}(z,u)|_{u=1}}{[z^{N_t}]C_t(z)},
\end{equation*}
and, making use of Eq. \ref{eq:denom}, we find
\begin{align}
    &\frac{\text{E}[n_m | N_t, I_t]}{I_t} = \frac{\binom{k_t+m-1}{m}\binom{k_t(I_t-1)+N_t-m-1}{N_t-m}}{\binom{k_tI_t+N_t-1}{N_t}} \label{eq:exact_pr}\\
    &\approx \binom{k_t+m-1}{m}p_t^{m}(1-p_t)^{k_t}\left[1 + \mathcal{O}\left(\frac{m}{I_t}\right)\right], \label{eq:prob}
\end{align}
where the first line is exact for a given trajectory, and the second line uses Sterling's approximation to the Gamma function in the form $\Gamma(x+\alpha) \sim \Gamma(x)x^{\alpha}[1 + (\alpha^2 + \alpha/2 + 1/12)/x]$ for $\alpha << x$. Eq. \ref{eq:prob} shows that the expected degree distribution of infectious nodes is asymptotically aligned with Eq. \ref{eq:NB} but perturbed by tree interactions that decay with $I_t$. 

We can prove further that the degree distribution converges to the expected distribution as graph size grows. A similar calculation, this time leveraging $C_{t,m}(z,u)$ to compute $\text{E}[n_m(n_m-1)|N_t,I_t]$, leads to an intuitive asymptotic variance estimate
\begin{align*}
    \text{V}[n_m|N_t,I_t] \approx I_tp(T_t=m)\left[1 - p(T_t = m)\right],
\end{align*}
implying that deviations around the expected $n_m$ are $\mathcal{O}(1/\sqrt{I_t})$, vanishing as $I_t$ grows. Thus, the infectious node degree distribution converges to Eq. \ref{eq:exact_pr} at a square-root rate, and $P(\ell|N_t,I_t) \approx \text{E}[n_{\ell} | N_t, I_t]/I_t$ is an approximation that improves with constellation size. In other words, as disease prevalence increases, we expect tree interactions to fade and constellation geometry to stabilize.

\begin{figure*}
\centering\includegraphics[width=\linewidth]{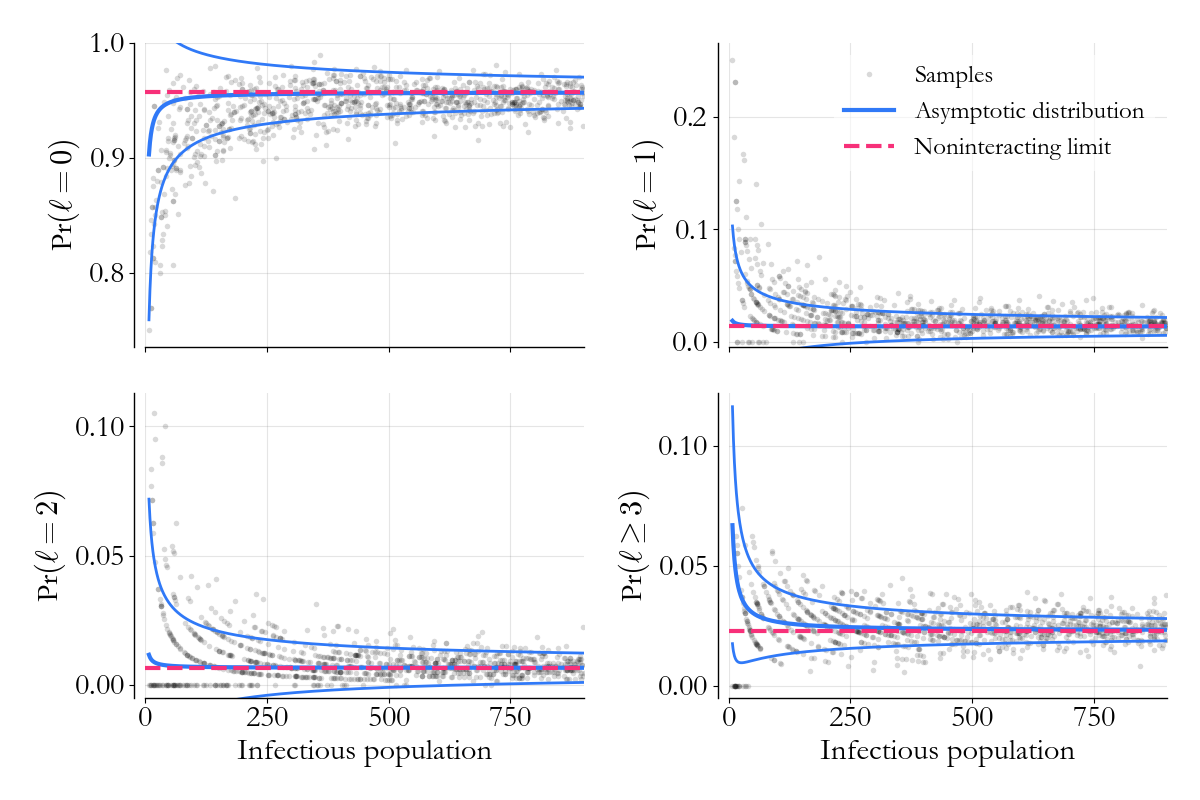}
\caption{Interactions and stability. Empirical degree distributions from 1000 sampled constellations (black dots) have interpretable dependence on the infectious node population. Generating function calculations (blue lines, mean and 2 standard deviations around it) show that the degree distribution converges to its expected value with negligible tree interactions (red dashed line) as disease prevalence increases.}
\label{fig:constellations}
\end{figure*}

We can validate these results by comparing $P(\ell |N_t,I_t)$ to sampled constellations of various size. This is shown in Fig. \ref{fig:constellations} for time-averaged values of $\mu_t$ and $k_t$ from the Washington COVID-19 model \cite{thakkar2022covid}. In the figure, dots come from the degree distribution across 1000 sampled graphs while the blue lines show the interacting generating function results (the mean and 2 standard deviations around it). At low $I_t$, tree interactions are significant and deviations from Eq. \ref{eq:NB} (red dashed line) are clear in both the samples and the generating function estimates. Meanwhile, volatility around the mean, both in the samples and in the asymptotic distribution decay with $I_t$. 

Fig. \ref{fig:constellations} illustrates the regime where tree interactions warp the epidemiology. This is a key structural feature of the transmission forest, particularly relevant in near elimination contexts, and it's something we intend to study further. But in the Washington COVID-19 model, from March 2020 to March 2021, the minimum $I_t$ is roughly 1500, implying that for the bulk of the model time period trees grow in functional isolation.

\section{Trees as recursive functions}

Moving from constellations to transmission trees requires us to link $P(\ell |N_t,I_t)$-distributed events over time, being careful to track when nodes start and stop being infectious. As we'll see, this process lends itself to a family of recursively defined generating functions that can be efficiently evaluated and analysed in reverse time.

The previous section's results can be written concisely in terms of the generating function for stars,
\begin{align*}
    s_t(z) &= \sum_{\ell \geq 0} P(\ell|N_t,I_t)z^{\ell}\\ 
    &\approx \sum_{\ell \geq 0} \frac{\text{E}[n_{\ell} | N_t, I_t]}{I_t} z^{\ell}\\
    &\approx \left(\frac{1-p_t}{1-p_tz}\right)^{k_t},
\end{align*}
where the first approximation comes from neglecting volatility around the average constellation, and the second approximation comes from neglecting tree interactions -- both valid above low prevalence. It's notable that this likely would've been our naive choice based on Eq. \ref{eq:NB} alone, but it's nice to have sound theoretical footing and a deeper understanding. 

In any case, for a given trajectory, we can formally define the set of all possible transmission trees rooted at time $r$, $\mathbb{T}_r$, by associating with a tree a monomial $\tau = z_{1}^{n_1}z_{2}^{n_2}\dotsm z_{\mathcal{T}}^{n_{\mathcal{T}}}$ where $n_t$ is the number of nodes at time $t$ up to final time $\mathcal{T}$. The set is finite since $0 \leq n_t \leq N_t$ for all $t$, and we can consider the generating function of complex vector $\mathbf{z} = (z_1,...,z_{\mathcal{T}})$
\begin{equation}
    T_{r}(\mathbf{z}) = \sum_{\tau \in \mathbb{T}_{r}} p(\tau|N_t,I_t) z_{r}z_{r+1}^{n_{r+1}}\dotsm z_{\mathcal{T}}^{n_{\mathcal{T}}},\label{eq:ftgf}
\end{equation}
which encapsulates tree structures rooted at time $r$. This expression is more of a formal statement than anything else since it's not clear how sums over $\mathbb{T}_{r}$ are executed, and $\mathbf{z}$ is $\mathcal{T}$-dimensional. We still need a more practical method for function evaluation. Towards that end, we define $\rho(t)$ as the probability of being infectious $t$ days after infection, which in the deterministic case is
\begin{equation*}
    \rho(t) = 
    \begin{cases} 
      1 & d_E < t \leq d_E+d_I \\
      0 & \text{otherwise},
   \end{cases}
\end{equation*}
but clearly could be defined based on a more sophisticated pathogenesis model. Then, for a binary process, the binomial generating function $b_t(z) = 1 - \rho(t) + \rho(t)z$ represents infectious status. 

With this book-keeping machinery in hand, we can make progress by recognizing that all trees are star graphs where the internal node is the root and the external nodes are replaced by trees. This is a well-known recursive idea \cite{flajolet2009analytic}, and it implies in our case that
\begin{equation}
    T_{r}(\mathbf{z}) = z_{r}\prod_{t\geq{1}}^{\mathcal{T}}\left[ 1 - \rho(t-r)+ \rho(t-r)s_{t}(T_{t}(\mathbf{z}))\right], \label{eq:tgf}
\end{equation}
linking root $z_r$ to new trees at the appropriate infectious times through nested compositions of $s_t(z)$ and $b_t(z)$.\footnote{Note that an additional function composition can be used to incorporate surveillance. If cases are reported with probability $\pi_t$, the binomial generating function $o_t(z) = 1 - \pi_t + \pi_t z$ determines if a node is observed, and we can replace $z_r$ with $o_{r}(z_r)$ directly.} The product in Eq. \ref{eq:tgf} is tractable given the boundary condition $T_{\mathcal{T}}(\mathbf{z}) = z_{\mathcal{T}}$, which captures the idea that a tree rooted at $\mathcal{T}$ has no time to grow (that is $\rho(t-\mathcal{T}) = 0$ for all $t$ under consideration). 

Eq. \ref{eq:tgf} defines a family of transmission tree generating functions indexed by their root times, but it doesn't give us a closed form expression for $T_r(\mathbf{z})$. In practice, Eq. \ref{eq:ftgf} reminds us of $T_r(\mathbf{z})$'s power series structure, and it motivates statistically meaningful function evaluations. Those evaluations can then be carried out in bespoke recursive programs based on Eq. \ref{eq:tgf}, starting at time $\mathcal{T}$ and working backwards.

\section{Calculating forest statistics}

To illustrate the theory's application and to simultaneously validate the sampling and recursive approaches, we take up 3 example calculations in this section: the size distribution of trees in the forest, the tree survival function, and the relationship between size and lifetime.

\textbf{Tree size.} Evaluating $T_{r}(\mathbf{z})$ at $\mathbf{z} = (z,z,...,z)$, that is dropping all delineation between $z_t$ over time, organizes trees by the number of nodes. Specifically, Eq. \ref{eq:ftgf} becomes
\begin{align*}
    T_r(z) &= \sum_{\tau\in\mathbb{T}_r} p(\tau | N_t, I_t) z^{n_r+n_{r+1}+\dotsm+n_{\mathcal{T}}}\\
    &= \sum_{n \geq 0} p(n|N_t,I_t) z^{n},
\end{align*}
where $n=\sum_{i=1}^{\mathcal{T}}n_i$, and in the second line we've grouped terms in the sum. Our goal is to calculate the coefficients in this now one-dimensional power series. 

Complex analysis offers an efficient and well-known path forward \cite{flajolet2009analytic}. For any analytic generating function $f(z)$, Cauchy's integral theorem implies that
\begin{equation*}
    [z^n]f(z) = \frac{1}{2\pi i} \oint \frac{f(\xi)}{\xi^{n+1}}d\xi,
\end{equation*}
over any closed contour in the complex plane. Choosing $|\xi|=1$ as the contour leads to 
\begin{align*}
    [z^n]f(z) &= \frac{1}{2\pi}\int_0^{2\pi} f\left(e^{i\theta}\right) e^{i n\theta}d\theta \\
    &\approx \frac{1}{M} \sum_{m=1}^M f\left(e^{2 \pi i \frac{m}{M}}\right)e^{2\pi i \frac{n m}{M}},
\end{align*}
for angle $\theta$, which we've discretized into $M$ points along the unit circle in the second line. The sum above is the discrete Fourier transform of $f(z)$, which implies that a set of coefficients can be extracted by passing the unit circle through $f(z)$ and transforming. Eq. \ref{eq:tgf} can be used to evaluate $T_{r}(z)$ on the unit circle directly, building function evaluations from $\mathcal{T}$ backwards.

\begin{figure*}
\centering\includegraphics[width=\linewidth]{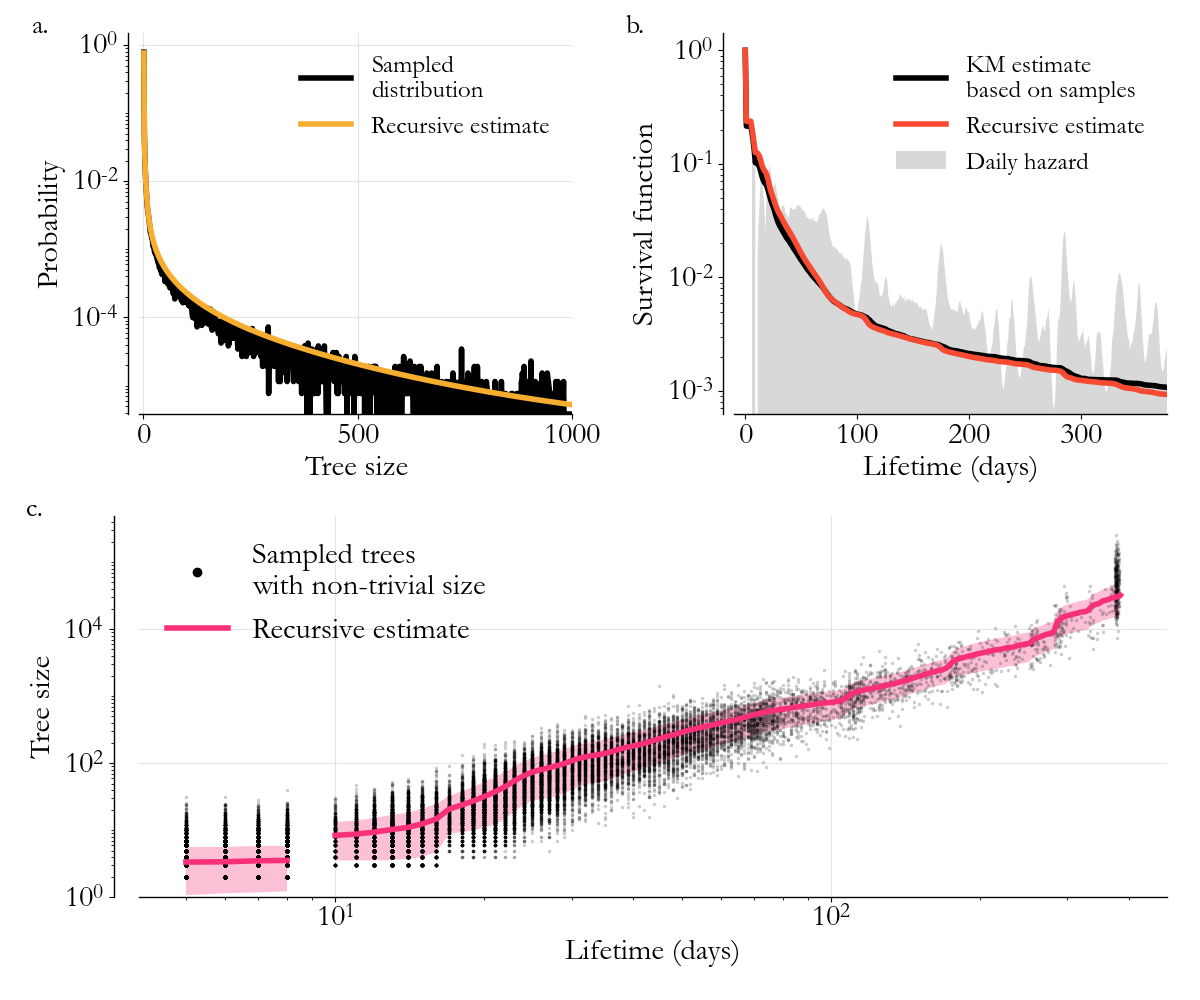}
\caption{Calculating forest statistics. (a) Complex-analysis based methods can be used to compute the size distribution of trees (yellow) which compares favorably to samples (black) (b) Structured evaluations can be used to compute the tree survival function (orange) which captures key features of a Kaplan-Meier estimator applied to samples (black) and can be used to compute the daily hazard function (grey). (c) Expectations (pink curve) and intervals (pink region) can be computed via Eq. \ref{eq:tgf} and its derivatives, shown here for the relationship between tree size and lifetime.}
\label{fig:vs_samples}
\end{figure*}

The results of this process are compared to 250k sample trees rooted in early March 2020 in Fig. \ref{fig:vs_samples}a, using the parameters from the Washington COVID-19 model \cite{thakkar2022covid}. Both methods yield consistent estimates, but with the generating function approach (yellow) remaining stable at very low probabilities. As a consequence of the skewed, individual-level transmission distribution (Eq. \ref{eq:NB}), tree size is heavy-tailed. Most trees are simply roots with little growth, but a small set of trees capture an out-sized fraction of nodes. In other words, intuitively, super spreaders imply the existence of super trees.

\textbf{Tree survival.} Spontaneous transmission tree extinction implies, in terms of monomials $\tau = z_{1}^{n_1}z_{2}^{n_2}\dotsm z_{\mathcal{T}}^{n_{\mathcal{T}}}$, that $n_t = 0$ for all $t$ above extinction time $t^*$. This observation motivates a systematic evaluation method for organizing transmission trees by their lifetimes. Consider $\mathbf{z}^* = (z_1, z_2, ..., z_{t^*}, 0, 0,...)$. Then,
\begin{equation*}
    T_{r}(\mathbf{z}^*) = \sum_{\tau \in \mathbb{T}_r} p(\tau,n_{t>t^{*}}=0|N_t,I_t) z_{r}z_{r+1}^{n_{r+1}}\dotsm z_{t^*}^{n_{t^*}},
\end{equation*}
which, upon setting all remaining $z_t=1$, gives
\begin{align*}
    T_{r}(\Theta(t^*)) &= \sum_{\tau \in \mathbb{T}_r} p(\tau,n_{t>t^{*}}=0|N_t,I_t) \\
    &= p(L \leq t^{*} - r)
\end{align*}
where $\Theta(t)$ is the Heaviside step function evaluated along $\mathbf{z}$ and $L$ is the tree lifetime. The survival function, $S_r(t^*) = P(L > t^* - r)$ is then
\begin{align}
    S_r(t^*) = 1 - T_{r}(\Theta(t^*)) \label{eq:survival}
\end{align}
which can be evaluated for all $t^*$ again using Eq. \ref{eq:tgf}. 

In Fig. \ref{fig:vs_samples}b, we compare Eq. \ref{eq:survival} to a Kaplan-Meier estimator applied to the same tree samples as above. The theory (orange) is consistent with the empirical approach (black), even in finer details. Small drops in survival correspond to times of more concerted transmission suppression in Washington, which are further highlighted in the hazard function (grey). We refer to past work \cite{thakkar2020social,thakkar2020one,thakkar2020towards} for epidemiological details, but these survival analysis concepts applied to transmission trees give perspective on intervention efficacy that complements more conventional measures like the effective reproductive number \cite{anderson1992infectious}. 

\textbf{Size vs. lifetime.} Finally, to illustrate how some statistics can be related, consider $\mathbf{z} \rightarrow u\mathbf{z}$ for complex $u$. Then, 
\begin{equation*}
    T_{r}(\mathbf{z}) = \sum_{\tau \in \mathbb{T}_{r}} p(\tau|N_t,I_t) z_{r}z_{r+1}^{n_{r+1}}\dotsm z_{\mathcal{T}}^{n_{\mathcal{T}}}u^n
\end{equation*}
consolidates tree size and structure. Based on the approach above, evaluations of the form
\begin{align*}
    &T_{r}(u\Theta(t^*)) - T_{r}(u\Theta(t^* - 1)) \\
    &= \sum_{\tau \in \mathbb{T}_r} p(\tau|L\leq t^* - r)u^n - \sum_{\tau \in \mathbb{T}_r} p(\tau|L\leq t^*-1- r)u^n\\
    &= \sum_{n\geq 0} p(n|L=t^*-r)u^n,
\end{align*}
can be used to relate tree size to the extinction time, and we can compute statistics. For example,
\begin{equation*}
    \text{E}[n|L=t^*-r] = \frac{\frac{\partial}{\partial u}[T_{r}(u\Theta(t^*)) - T_{r}(u\Theta(t^* - 1))]_{u=1}}{T_{r}(\Theta(t^*)) - T_{r}(\Theta(t^* - 1))},
\end{equation*}
is the expected size at a fixed lifetime. To evaluate the numerator, we can differentiate Eq. \ref{eq:tgf} to derive a recursion relation for $T_r(\mathbf{z})$'s partial derivatives. We find
\begin{widetext}
\begin{equation}
    \frac{\partial}{\partial u} T_{r}(u\mathbf{z}) = T_{r}(u\mathbf{z}) \left(\frac{1}{u} + \sum_{t=1}^{\mathcal{T}} \frac{\rho(t-r)s_t^{\prime}(T_t(u\mathbf{z}))}{1 - \rho(t-r) + \rho(t-r)s_t(T_t(u\mathbf{z}))}\frac{\partial}{\partial u} T_{t}(u\mathbf{z})\right) \label{eq:partial}
\end{equation}
\end{widetext}
where primes denote total derivatives and the evaluation depends on the full family of rooted generating functions $T_r(\mathbf{z})$. Practically, for some specific $\mathbf{z}$, we can evaluate all $T_{r}(\mathbf{z})$ first, which then specifies the details in Eq. \ref{eq:partial}. A similar calculation can be used to compute higher order derivatives and corresponding higher order statistics. 

We compare the generating function approach above to samples with non-zero lifetime in Fig. \ref{fig:vs_samples}c. Statistics based on Eq. \ref{eq:tgf} (pink) gracefully capture discretization effects, like the necessary spacing between trees surviving one and two generations, as well as the transition to continuous lifetimes. Features of the relationship, like local slope changes in keeping with the mitigation environment in Washington at that time, are clearly reflected in both the model and the samples. That this distribution can be calculated efficiently based on time series data alone is striking, and it illustrates how accessible transmission tree statistics might be with the right approach.

\section{Conclusion}

The key idea in this paper is that conventional stochastic process models of disease transmission, based on Eq. \ref{eq:sir_view}, can be used to estimate transmission tree properties without additional data. The bridge between the population-scale of trajectories and the individual-scale of trees comes from Eq. \ref{eq:bp_view}, which led us to a generating function family that can be used to efficiently compute a variety of epidemiologically relevant statistics. 

The theory developed here offers new perspective on individual-level data, like distributions across outbreak investigations or associations among collections of genetic sequences. Those types of comparisons were made empirically in our previous paper \cite{thakkar2022covid}, but the analytic approach gives additional paths towards more quantitative comparisons and potentially joint inferences. Speaking broadly, tree structures have a long mathematical history. Eq. \ref{eq:tgf} represents a promising and intuitive connection between that body of work and the classical compartmental models used to describe disease transmission.

\section*{Acknowledgements}
This work was done with many people's support. In particular, we want to thank Kevin McCarthy for his attention and detailed comments. His input made the writing much more clear and it brought to light implications we hadn't considered.

\bibliography{references}

\end{document}